\documentclass[preprint,12pt,authoryear]{elsarticle}

\usepackage{amssymb}
\usepackage{amsmath}

\usepackage[
    left=1in,
    right=1in,
    top=1in,
    bottom=1in
]{geometry}

\usepackage{orcidlink}
\usepackage{graphicx}%
\usepackage{multirow}%
\usepackage{amsmath,amssymb,amsfonts}%
\usepackage{amsthm}%
\usepackage{mathrsfs}%
\usepackage[title]{appendix}%
\usepackage{xcolor}%
\usepackage{textcomp}%
\usepackage{manyfoot}%
\usepackage{booktabs}%
\usepackage{algorithm}%
\usepackage{algorithmicx}%
\usepackage{algpseudocode}%
\usepackage{listings}%

\usepackage[T1]{fontenc}

\usepackage{hyperref}

\usepackage{subcaption}
\usepackage{tcolorbox}

\usepackage{array}
\usepackage{minted}

\newcommand{\spacedinlineheader}[1]{\vspace{2mm} \noindent {\em #1: }}

\journal{}

\begin{document}

\begin{frontmatter}

\title{Consistently Good vs. Occasionally Great: A Rubric for Open-Ended Feedback Quality from Humans and Machines}

\author[1]{Binglin Chen\,\orcidlink{0000-0001-9033-1281}}
\ead{chen386@illinois.edu}

\author[1]{Rajarshi Haldar\,\orcidlink{0000-0001-8464-3211}}
\ead{rhaldar2@illinois.edu}

\author[1]{Max Fowler\,\orcidlink{0000-0002-4730-447X}}
\ead{mfowler5@illinois.edu}

\author[1]{Matthew West\,\orcidlink{0000-0002-7605-0050}}
\ead{mwest@illinois.edu}

\author[1]{Craig Zilles\,\orcidlink{0000-0003-4601-4398}}
\ead{zilles@illinois.edu}

\affiliation[1]{
    organization={University of Illinois Urbana-Champaign},
    city={Urbana},
    state={IL},
    country={USA}
}

\begin{abstract}
Providing high-quality feedback on student work is essential for learning, yet delivering such feedback at scale remains challenging. In this paper, we focus on feedback for open-ended short answer questions in introductory programming, with the goal of nudging students toward success on reattempts without revealing the correct answer. We develop a five-criteria rubric grounded in educational literature for evaluating feedback quality: (1) acknowledging correct portions of the student answer, (2) identifying at least one flaw (if present), (3) providing actionable guidance for improvement, (4) maintaining appropriate concealment of the answer, and (5) using an appropriate conversational tone. Using this rubric, we compare feedback generated by a frontier LLM (OpenAI o1) to feedback from nine teaching assistants across 90 student responses, with three researchers and an LLM independently scoring all feedback. Our results show that while one TA often produced the best feedback, the LLM demonstrated consistently higher average performance than TAs, as evaluated by humans. However, we also uncover significant self-preference bias when using LLMs to evaluate feedback quality: the LLM systematically rated its own outputs higher than human experts did. This bias, which research suggests persists even in cross-model evaluation, raises important methodological concerns for researchers employing LLM-based evaluation. We provide detailed characterization of both TA and LLM performance, analyze sources of variance in TA feedback quality, and discuss implications for deploying LLM-generated feedback in educational settings.
\end{abstract}

\begin{keyword}
Feedback evaluation \sep Human vs. GenAI \sep Short answer questions
\end{keyword}

\end{frontmatter}

\section{Introduction}

Feedback on student work is a crucial component of education~\cite{schartel2012giving}. However, delivering high-quality feedback in real time at scale is impractical for human instructors. Before large language models (LLMs), efforts to automate feedback often utilized structured question formats, such as multiple-choice and programming questions to provide predefined feedback messages, while other approaches attempted to train classifiers on past student responses to select predefined feedback messages, but lacked the ability to provide open-ended, text-based feedback~\cite{cavalcanti2021automatic,hahn2021systematic}. The trajectory of LLMs, particularly in reasoning capabilities, suggests that providing open-ended feedback for open-ended questions is now within reach.

\begin{figure}[t]
    \centering
    \includegraphics[width=.95\columnwidth]{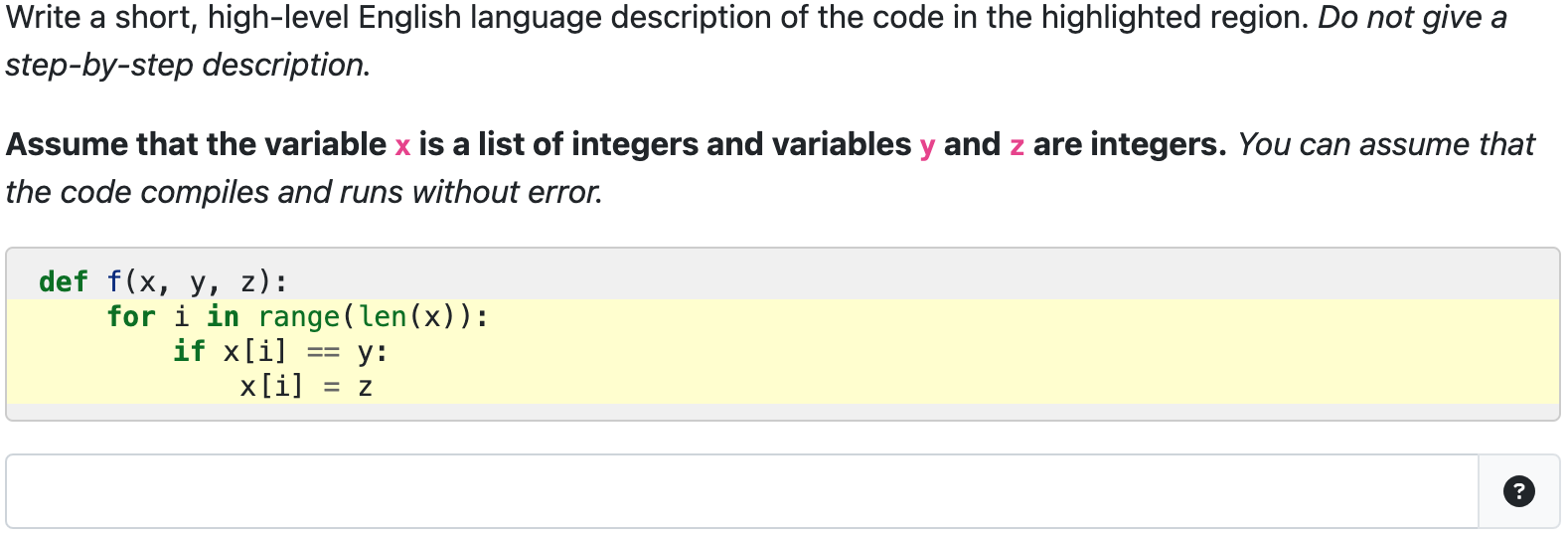}
    \caption{Example Explain in Plain English question.}
    \label{fig:code_reading_question}
\end{figure}

We envision GenAI systems that interactively guide students to solve open-ended questions by giving feedback that nudges them towards the answer. For early (incorrect) responses to each question, it is important to avoid giving away the answer so the student has the opportunity to solve the question themselves. As the student repeatedly attempts the question, the GenAI should provide increasingly large hints until the answer is eventually revealed, providing the struggling student with a worked example. To gauge our progress towards this vision, we propose a rubric (Section~\ref{sec:rubric}) to evaluate feedback quality for open-ended short answer questions, focusing on the very first interaction between the GenAI and student. This rubric is motivated by foundational literature in feedback (Section~\ref{sec:related_work}) and refined by reviewing feedback generated from a broad range of LLM models. The short answer questions we use in this paper are Explain in Plain English (EiPE) prompts~\cite{murphy2012ability}, where the student is asked to describe (in plain English) the purpose of a code snippet (see Figure~\ref{fig:code_reading_question} for an example EiPE question).

We explore three research questions through rating 90 feedback messages from a frontier LLM and 270 from nine teaching assistants (TAs) (Section~\ref{sec:data_collection}), as evaluated by three authors (Section~\ref{sec:evaluation_process}):
\begin{itemize}
    \item RQ1: What characterizes high-quality feedback for open-ended short answer questions, and can it be reliably assessed?
    \item RQ2: How does the quality of LLM-generated feedback compare to TA-generated feedback?
    \item RQ3: How does LLM evaluation of feedback quality compare to human expert evaluation?
\end{itemize}

Our results (Section~\ref{sec:results}) demonstrate that LLM-generated feedback is comparable to, if not better than, TA-generated feedback across multiple quality criteria. However, we also uncover significant self-preference bias in LLM evaluation, where LLMs systematically rate their own outputs higher than human experts do---a pattern that research suggests persists even when using different models as evaluators. We discuss the implications of these findings (Section~\ref{sec:discussion}), including reflections on the rubric, the comparative quality of LLM and TA feedback, and methodological concerns with LLM-based evaluation.

\section{Related Work}
\label{sec:related_work}

In this section, we review the literature on effective feedback practices and the emerging role of LLMs in automated feedback generation. We begin by examining foundational principles of effective feedback established in educational research. We then review recent empirical studies evaluating LLMs as feedback providers, including randomized controlled trials measuring learning outcomes and direct evaluations of feedback quality. Next, we explore the use of LLMs as automated feedback evaluators, a critical consideration for scaling feedback quality assessment. Finally, we provide context on Explain in Plain English (EiPE) questions, the specific task domain for this study.

\subsection{Foundational insights on effective feedback}

In perhaps one of the most influential contributions to the feedback literature, Hattie~\cite{hattie2007power} proposed that effective feedback should address three key questions: ``Where am I going?'' (What are the goals?), ``How am I going?'' (What progress is being made toward the goal?), and ``Where to next?'' (What activities need to be undertaken to make better progress?) Although the first question may not always be central in automatic feedback generation
, the latter two questions carry significant implications for what constitutes high-quality feedback.

Around the same time, Shute~\cite{shute2008focus} reviewed the literature on formative feedback and documented conclusions that may now seem self-evident. For example, effective feedback should include both verifying whether an answer is correct and elaborating on the correct answer or why the given answer is wrong. It should also be specific---excessive complexity or length can overwhelm students---and constructive, guiding students to re-attempt the problem without giving away the answer. These observations remain central to contemporary perspectives on delivering effective feedback.

In a parallel line of research, Mayer and colleagues explored ways to present instructional text in multi-media learning and found that using conversational style rather than formal style (e.g. ``your lung'' instead of ``the lung'') improved students' retention and transfer~\cite{moreno2000engaging, moreno2004personalized, mayer2004personalization}. This was later confirmed with a meta-analysis by Ginns et al.~\cite{ginns2013designing}. These findings on the effects of conversational language suggest that feedback delivered in a conversational tone may be more effective than formal, impersonal language.

\subsection{LLMs as feedback providers, randomized controlled trials}


Pankiewicz and Baker conducted a randomized controlled trial with 248 students at a European university investigating GPT-4o-generated feedback on compiler errors in an introductory programming course. Students in the experimental group received LLM feedback in addition to standard feedback (compiler messages and unit test results when applicable), while the control group received only standard feedback. Feedback usefulness was rated by students in the experimental group via a 5-point Likert scale pop-up (appearing after every fifth feedback instance), affective states for both groups were collected via self-report pop-ups (appearing randomly with 1/10 probability), and performance was measured on a pre-test and post-test that were two months apart. Pankiewicz and Baker reported that 81\% of students in the experimental group rated the feedback as very or extremely useful (median rating = 4.5 out of 5). Affective surveys revealed that the experimental group self-reported higher focus and lower confusion and frustration after compiler errors. When GPT feedback was disabled during the post-test, the experimental group completed assignments significantly faster (27min vs. 33min, p=.007) and showed faster error resolution. However, no significant differences were observed in post-test scores between groups ($\beta$=-0.082, p=.427)~\cite{pankiewicz2025enhancing}.

Meyer et al. conducted a randomized controlled trial with 459 upper secondary students (Grade 10) learning English as a foreign language, who were tasked with writing argumentative essays for TOEFL iBT tasks. Students in the experimental group received feedback generated by GPT-3.5-turbo and were asked to revise their texts accordingly, while the control group revised their essays without receiving any feedback (only a general instruction to revise). Students then completed a second writing task on a different topic. An automated essay scoring algorithm trained on expert-rated texts (achieving a quadratic weighted kappa of 0.76 with human experts) assessed students' writing performance across all tasks. Meyer et al. reported that LLM-generated feedback significantly improved revision performance with a small effect size (Cohen's d = 0.19) compared to no feedback. However, no significant differences were observed in performance on the subsequent new writing task (Cohen's d = 0.13, p = .259). Additionally, students who received LLM-generated feedback reported significantly higher task motivation for future writing tasks (Cohen's d = 0.36), experienced more positive emotions (Cohen's d = 0.34), and rated the feedback as significantly more useful (Cohen's d = 1.19) compared to the control group's general revision instruction~\cite{meyer2024using}.


Kinder et al. conducted a randomized controlled trial with 269 master's-level German pre-service teachers, who were tasked with reasoning about a pedagogical recommendation made by a fictitious teacher for a fictitious student in a case scenario. Participants wrote a paragraph stating whether they agreed with the teacher's diagnosis and justifying their decision. The experimental group received adaptive feedback generated by GPT-4 tailored to each individual response, while the control group received static expert-written feedback that included the correct decision and explanations of several ways to justify it, which was identical for all students in that condition. Participants were then presented with a new scenario and asked to write another evaluation. Two human raters blindly assessed students' writing on this second task for decision accuracy and justification quality, achieving high inter-rater reliability (Cohen's Kappa = 0.88). Kinder et al. reported that adaptive feedback significantly improved justification quality with a small effect size (Cohen's $f^2$ = 0.017) compared to static feedback. However, no significant difference was observed in decision accuracy on the new scenario between the two groups. Additionally, students who received adaptive feedback wrote longer texts in the subsequent task (medium effect size, Cohen's $f^2$ = 0.170), rated the feedback as more useful and interesting, and spent more time processing it per word~\cite{kinder2025effects}.


Zhou et al. investigated how LLM-generated feedback for compiler errors impacts learner persistence in programming tasks through a randomized controlled trial with 257 CS1 students at a large European university. Students were randomly assigned to either receive standard compiler error messages plus GPT-4-generated feedback (experimental group, N=129) or standard compiler error messages only (control group, N=128) across 141 C\# programming tasks. For the first half of each module (72 tasks), the experimental group received LLM feedback, which was then withdrawn for the latter half of each module (69 tasks) to assess lasting effects. Zhou et al. reported that with LLM feedback enabled, the experimental group showed significantly higher last submission scores ($\beta$=0.029, p=0.041), marginally higher problem-solving rates ($\beta$=0.028, p=0.061), and significantly lower rates of exceeding time thresholds ($\beta$=-0.098, p=0.002), taking breaks ($\beta$=-0.035, p=0.050), and wheel spinning ($\beta$=-0.025, p=0.032). Similar benefits were observed for specific challenging tasks. However, when LLM feedback was removed, almost all observed benefits disappeared, with no significant differences between the groups~\cite{zhou2025impact}.


Rüdian et al. compared students' perceptions of human feedback and LLM-generated feedback in a teacher education seminar for elementary school teaching at a German university. They conducted a study with 32 students who submitted essays on language teaching. Students were randomly assigned to receive either manually created teacher feedback or few-shot LLM-generated feedback (using Llama 3.0 70B prompted with 10 historical student submissions and corresponding teacher feedback). Students were then surveyed about the feedback they received. Results showed that when students were unaware of the feedback source, no significant differences emerged in perceived helpfulness or motivational aspects between teacher-created and LLM-generated feedback. Students were unable to reliably identify the feedback source (only 13 out of 22 students who made a guess identified it correctly, 59\% accuracy)~\cite{rudian2025feedback}.

\subsection{LLMs as feedback providers, direct evaluation}


Estévez-Ayres et al. evaluated the ability of LLMs to provide feedback on concurrent programming exercises. Using 52 student submissions in Java from a systems architecture course at a Spanish university, they experimented with GPT-3.5 and PaLM 2 using prompts of different levels of context, specifically focusing on their ability to detect three typical synchronization errors (deadlock, race condition, and starvation). The course instructor graded all 52 submissions and identified that 10 submissions (19\%) contained deadlocks, 11 (21\%) contained race conditions, 2 (4\%) contained starvation, and 35 (67\%) were error-free. Estévez-Ayres et al. reported that averaged across all prompts and synchronization errors, both models achieved approximately 50\% accuracy in detecting synchronization errors. Although recall values were often close to 1.0, precision remained consistently low (highest: 0.44), indicating that the models were overly risk-averse and frequently identified errors where none existed. The best performance was achieved by PaLM 2 in detecting race conditions (accuracy: 0.74, F1: 0.59)~\cite{estevez2025evaluation}.

Dai et al. investigated the proficiency of GPT models in generating assessment feedback on student writing in a postgraduate data science course. Using 103 student project proposals, they prompted GPT-3.5 and GPT-4 to generate feedback on five assessment aspects (goal description, topic appropriateness, business benefits, novelty/creativity, and overall clarity) specified in the marking rubric, and compared the generated feedback with corresponding instructor feedback. Two experts annotated all generated feedback on whether they agreed with instructor feedback polarity (positive, negative, or not mentioned) on each of the five assessment aspects, achieving high inter-rater agreement (Cohen's kappa > 0.8 across the board). Dai et al. reported that GPT-4 achieved higher F1 scores than GPT-3.5 with respect to human instructor feedback across most assessment aspects, with the highest agreement with human instructor feedback on the topic appropriateness aspect (F1 = 0.91). However, both models showed low reliability on other aspects (F1 < 0.60). Notably, since instructors often did not comment on all aspects, Dai et al. also evaluated GPT-4's feedback on aspects the instructor had not mentioned, finding that 90.10\% of these additional comments received approval from a human assessor. Two experts also annotated all feedback using Hattie and Timperley's framework~\cite{hattie2007power} (feeding-up, feeding-back, feeding-forward; task, process, self-regulation, and self levels), achieving high inter-rater agreement (Cohen's kappa > 0.7 across the board). Dai et al. reported that GPT-4 outperformed both GPT-3.5 and human instructors in consistently providing the aforementioned feedback dimensions, particularly feeding-up (GPT-4: 94.17\%, GPT-3.5: 0\%, human instructors: 5.83\%) and process-level feedback (GPT-4: 97.09\%, GPT-3.5: 55\%, human instructors: 80\%)~\cite{dai2024assessing}.

Stahl et al. experimented with LLM-based prompting strategies for automated essay scoring (AES) and feedback generation on student essays from grades 7-10 using the ASAP dataset and Mistral-7B-Instruct-v0.2. A key comparison was between joint approaches (generating both scores and feedback together) versus feedback-only generation (producing only feedback without scoring). Feedback quality was evaluated both automatically (using Mistral and Llama-2 to rate feedback on a 1-10 scale) and manually (twelve annotators rated 24 texts on a 7-point Likert scale across five dimensions: clarity, error explanation, precision, age-appropriateness, and overall helpfulness). Results showed that feedback-only generation outperformed joint approaches (mean 5.89/7 vs. lower scores for joint methods), though all scored above neutral (4/7), suggesting that generated feedback does not necessarily benefit from joint task formulation. For automatic evaluation, Mistral showed moderate correlation with human judgments (mean r = 0.37 with highest r = 0.61 for the overall helpfulness dimension), while Llama-2 showed negligible correlations across all dimensions, indicating that LLM-based automatic evaluation shows some promise but requires careful model selection~\cite{stahl2024exploring}.

Balse et al. evaluated the quality of LLM-generated explanations for logical errors in CS1 student programs through TA ranking and expert tagging. Using 30 buggy student solutions across 6 Python programming problems from an online CS1 exam, they compared explanations generated by GPT-3.5-turbo with those created by five undergraduate TAs. In a blind ranking task, TAs ranked LLM explanations comparably to peer-generated explanations, with the LLM achieving an average rank of 2.48 compared to the top-performing TA's 2.45. Through manual tagging by two experienced CS instructors, they found that 28/30 (93\%) of LLM-generated explanations identified at least one logical error, and 10/30 (33\%) were completely correct with no inaccuracies or omissions. However, 15/30 (50\%) contained at least one incorrect statement, and 10/30 (33\%) were missing at least one logical error~\cite{balse2023evaluating}.


While not the main focus of their work, Scarlatos et al. proposed a rubric to evaluate feedback on middle school math multiple-choice questions, using five binary-scale rubric dimensions: (1) does not make any incorrect statements, (2) does not reveal the answer, (3) provides a good suggestion, (4) points out errors, and (5) maintains a positive tone. They applied this rubric to feedback generated by both GPT-4 and teachers, finding that GPT-4 achieved near-perfect scores on every dimension and outperformed teachers by a large margin on two of them. However, they suspected that GPT-4’s high performance might be inflated due to the use of GPT-4 itself as the primary evaluator, potentially favoring AI-generated feedback despite a 76\% average agreement between GPT-4 and human evaluators. Additionally, teachers who provided feedback were unaware of the rubric criteria when writing, which may have contributed to the observed performance gap~\cite{scarlatos2024improving}.

Steiss et al. evaluated essay feedback using a rubric with five dimensions, each on a 5-point scale: (1) is criteria-based, (2) provides clear directions for improvement, (3) is accurate, (4) prioritizes essential features, and (5) uses a supportive tone. They compared feedback provided on 200 student essays (Grades 6-12) by GPT-3.5 and 16 human educators, 12 of whom had over 15 years of experience teaching writing. All human educators also received three hours of training in effective feedback practices. The results showed that humans outperformed GPT-3.5 on all dimensions except the first dimension listed above, where humans performed worse, but the differences were modest~\cite{steiss2024comparing}.

Wan and Chen explored the use of GPT-3.5 to generate feedback on students' responses to a single physics conceptual question. Four student researchers evaluated 16 pairs of human and AI-generated feedback on student responses, rating them equally in correctness, but finding the AI-generated feedback more helpful. This may be because their human feedback tended to be shorter than GPT-3.5's feedback. Additionally, four instructors assessed GPT-3.5's feedback on 65 student responses using a four-point scale, finding that approximately 70\% required only minor or no modifications before being delivered to students~\cite{wan2024exploring}.


\subsection{LLMs as feedback evaluators}

Koutcheme et al. investigated whether open-source LLMs can generate quality programming feedback and whether GPT-4 can reliably evaluate such feedback. Using a dataset of 150 help requests from an introductory Dart programming course with accompanying GPT-3.5-generated feedback previously evaluated by human experts, they first assessed GPT-4's ability to judge feedback quality by comparing its evaluations against human expert ratings across three criteria: completeness (identifying all issues), perceptivity (identifying at least one issue), and selectivity (avoiding hallucinated issues). GPT-4 demonstrated moderate agreement with human raters (Cohen's kappa of 0.48 for completeness, 0.22 for perceptivity, and 0.40 for selectivity), although it exhibited a positive bias, rating feedback more favorably than human experts. Koutcheme et al. then used GPT-4 as an automated judge to evaluate the feedback generated by several open-source models (from the CodeLlama and Zephyr families) alongside proprietary models (GPT-3.5 and GPT-4). Results showed that smaller open-source models like Zephyr-7B-$\beta$ performed comparably to GPT-3.5, with both models producing feedback that met all three criteria in approximately 70\% of cases, compared to 99\% for GPT-4~\cite{koutcheme2024open}.

Jia et al. investigated hallucinations in LLM-generated feedback on student project reports by comparing data-driven approaches (fine-tuning a BART model) and prompt-driven approaches (prompting GPT-4 directly), as well as methods for automatically detecting such hallucinations. Using a dataset of 484 student project reports with instructor feedback from a graduate-level object-oriented development course, they trained the BART model on 434 report-feedback pairs and generated feedback for 50 test reports. Two human annotators evaluated the generated feedback for hallucinations, categorizing whether each feedback sentence contained hallucinated content, achieving substantial inter-annotator agreement (Cohen's kappa = 0.693). Jia et al. reported that both approaches generated substantial hallucinated content: 27.1\% for the data-driven approach and 23.5\% for the prompt-driven approach. For automatic hallucination detection, they found that ChatGPT-3.5-based methods performed better than NLI-based methods overall, and instruction fine-tuning improved both methods by approximately 0.12 F1-wise, with fine-tuned ChatGPT-3.5 ultimately achieving an F1 of 0.572 for detecting hallucinations~\cite{jia2024assessing}.

While not primarily focused on assessing LLMs' evaluation capabilities, Scarlatos et al. explored using LLMs as the primary evaluator to obtain training signals for DPO. They trained Llama 3.1 8B Instruct to generate effective tutor utterances using two automated evaluation methods: LLMKT (predicting student correctness, 0.76 AUC) and GPT-4o (assessing pedagogical quality on six rubric criteria plus an overall 1-10 score). The DPO-trained model significantly outperformed baselines (including GPT-4o) on LLMKT-based predictions  while achieving comparable GPT-4o-based pedagogical scores. Human evaluation with two annotators on 50 instances confirmed better performance. However, inter-rater reliability was low: human-human agreement showed Kendall's $\tau$ = 0.06 for correctness rankings and  Pearson's $\rho$ = 0.15 for rubric scores, while human-LLM agreement averaged $\tau$ = 0.08 for correctness rankings (with LLMKT) and $\rho$ = 0.27 for rubric scores (with GPT-4o)~\cite{scarlatos2025training}.

\subsection{Explain in plain English (EipE) questions}

The Explain in Plain English (EiPE) task used in this paper has recently attracted interest due to its relationship to prompting LLMs to generate code. 
Concurrent with broader research exploring the implications of LLMs on computing education~\cite{kiesler2023large,prather2023robots,prather_2024_wg,reeves2023evaluating}, there has been significant effort in the use of LLMs as a grader for EiPE tasks, as well as investigating the relationship between the EiPE task and the skill of prompting~\cite{kerslake2024integrating,prather2024breaking,smith2024prompting,smith2024explain,smith2024code,smith2024evaluating}, 
including the introduction of explicit prompting problems~\cite{denny2023prompt}.

\section{The Rubric}
\label{sec:rubric}

Our rubric consists of five criteria, each rated on a 3-point scale (0--2). An overall score is computed as the sum of all five criteria scores, ranging from 0 to 10. Below, we provide an overview of what each criterion aims to capture and the rationale behind it. The complete rubric description can be found in Figure~\ref{fig:rubric_detail}.


\spacedinlineheader{Acknowledgment of Understanding}Good feedback acknowledges what the student has done correctly, when applicable, to reinforce their understanding.

As Hattie posits, effective feedback should answer the question ``How am I going?''~\cite{hattie2007power} Informing students about what they have done correctly is therefore a crucial aspect of feedback. Anecdotally, we have observed that when students receive only binary correct/incorrect feedback, they sometimes modify parts of their answers that were actually correct, ultimately moving further away from a correct solution.

\spacedinlineheader{Identification of Issues}Good feedback highlights key errors, misunderstandings, and omissions of important details exhibited in the response.

Similarly to Acknowledgment, or rather, as two sides of the same coin, Acknowledgment of Understanding and Identification of Issues together address Hattie's ``How am I going?'' question~\cite{hattie2007power}. While Acknowledgment reinforces what students have done correctly, Identification ensures they understand where their response falls short. Both are essential for providing balanced and effective feedback that guides students toward meaningful learning and improvement.

\spacedinlineheader{Guidance for Improvements}Good feedback offers students guidance that would improve their answer or their understanding.

As Hattie suggests, effective feedback also needs to address the question ``Where to next?''~\cite{hattie2007power} Without proper guidance, students may struggle to determine their next steps, leading to frustration or unproductive trial-and-error revisions. Feedback should help students progress toward their learning goals by offering clear direction---whether through thought-provoking questions, key concepts to review, alternative approaches to consider, or even an explanation or walk-through of the correct answer, as long as they still have new questions to practice and reinforce their learning.

\spacedinlineheader{Concealment of Answers}Good feedback avoids revealing the correct answer beyond what the student already understands to promote independent thinking and learning.

In her seminal review, Shute distinguished between facilitative feedback (cues) and directive feedback (corrective information), suggesting that facilitative feedback enhances learning, provided that students have sufficient mastery of the topic~\cite{shute2008focus}. Based on this, we posit that initial feedback---which is the focus of this rubric---should prioritize a facilitative approach, meaning that correct answers should be concealed. Only when there is clear evidence that facilitative feedback is ineffective should feedback become more directive. This approach mirrors traditional intelligent tutoring systems, which gradually increase the scaffolding as the student remains stuck on a problem.

\spacedinlineheader{Conversationality of Language}Good feedback is written as if it is part of a natural, conversational exchange between a teacher or teaching assistant and the student and avoids language problems that may confuse or mislead the student.

In a series of experiments, Mayer and colleagues demonstrated that students learn better when instructional text is presented in conversational style rather than formal style (i.e. ``your answer'' instead of ``the answer'')~\cite{moreno2000engaging, moreno2004personalized, mayer2004personalization}. Based on this, we posit that feedback with a conversational tone is more likely to encourage students to actively engage with it, fostering a more supportive learning environment. While humans are naturally capable of providing conversational feedback, the same cannot always be said for GenAI. Smaller models, in particular, sometimes struggle to generate comprehensible feedback based on our past experiments with Llama 2 7B. Ensuring that AI-generated feedback maintains clarity and a natural conversational tone is essential to make it as effective as human-provided feedback.

\begin{figure}[!ht]
    \centering
    \begin{tcolorbox}[width=\textwidth, colback=gray!10, colframe=black]
    \textbf{Acknowledgment of Understanding}: \\
    Good feedback acknowledges what the student has done correctly, when applicable, to reinforce their understanding. \\
    
    {\bf 2 points:} The feedback acknowledges some important key correct elements of the response in a way that is meaningful and appropriate to the student's current understanding. This includes cases where the feedback says the student is close or captures the essence of the code and the student actually is one or two steps away. \\
    {\bf 1 point:} The feedback overemphasizes less important aspects, or uses vague, general phrases (e.g., ``Good job'') that fail to specify what the student did well. \\
    {\bf 0 points:} The feedback incorrectly praises incorrect/non-existent elements, fails to acknowledge any correct components when present, or gives unwarranted praise that doesn't match the student's understanding. \\ \\
    
    \textbf{Identification of Issues}: \\
    Good feedback highlights key errors, misunderstandings, and omissions of important details exhibited in the response. \\
    
    {\bf 2 points:} The feedback identifies and explicitly states one or more critical errors, misunderstandings, or omissions of important details clearly and specifically. If proposing something that is correct instead of something else that is not, it should be interpreted as explicit identification. \\
    {\bf 1 point:} The feedback identifies an issue but may lack clarity, depth, or focus on less significant problems. It may use vague or ambiguous language, that makes it harder for the student to pinpoint the real problem. Implicit identification also falls here (e.g., a suggestion implies an issue without explicitly stating it). \\
    {\bf 0 points:} The feedback either fails to identify an issue that exists, mislabels something correct as incorrect, or overlooks the key issue altogether. \\
    \end{tcolorbox}
    \caption{Complete rubric description (continues on next page)}
\end{figure}

\begin{figure}[!ht] \ContinuedFloat
    \centering
    \begin{tcolorbox}[width=\textwidth, colback=gray!10, colframe=black]
    \textbf{Guidance for Improvements}: \\
    Good feedback offers students guidance that would improve their answer or their understanding. \\
    
    {\bf 2 points:} The feedback provides clear, and specific guidance that is highly likely to improve the student's answer or deepen their understanding of the topic. Directly revealing or walking through a correct answer also falls here.  This can also include suggestions to remove a strictly wrong part that improves an answer. \\
    {\bf 1 point:} The feedback provides some guidance, but it is vague, only partially or indirectly addresses the issue, or doesn't fully align with the student's understanding. It may help, but it lacks precision or clear action steps. A mix of good and bad guidance also falls here. \\
    {\bf 0 points:} The feedback either lacks guidance, provides a misleading suggestion, or suggests changes that make the response worse. Generic or vague suggestions with no actionable value also fall here. \\ \\
    
    \textbf{Concealment of Answers}: \\
    Good feedback avoids revealing the correct answer beyond what the student already understands to promote independent thinking and learning. \\
    
    {\bf 2 points:} The feedback avoids revealing the answer/understanding entirely (beyond what the student already understands), instead providing hints, clarifications, or probing questions that encourage the student to reflect and arrive at the answer themselves. \\
    {\bf 1 point:} The feedback reveals parts of the answer beyond what the student already knows, such as key concepts or missing words. It may provide significant guidance or a low-level solution but stops short of handing over a fully copy-pastable answer. or the exact piece of the answer the student needs. \\
    {\bf 0 points:} The feedback directly gives the full answer, provides the exact missing piece, or includes an incorrect answer.
    \end{tcolorbox}
    \caption{Complete rubric description (continues on next page)}
\end{figure}

\begin{figure}[!ht] \ContinuedFloat
    \centering
    \begin{tcolorbox}[width=\textwidth, colback=gray!10, colframe=black]
    \textbf{Conversationality of Language}: \\
    Good feedback is written as if it is part of a natural, conversational exchange between a teacher or teaching assistant and the student, and avoids language problems that may confuse or mislead the student. \\
    
    {\bf 2 points:} The feedback is clear, concise, and grammatically correct, making it easy to understand. It's written naturally in a few sentences, speaking directly to the student without unnecessary formality or structure. It speaks directly to the student, avoiding third-person references or role-playing. It stays brief and natural, without lists, sections, or formalities like greetings and sign-offs. \\
    {\bf 1 point:} The feedback includes elements that make it feel less like a direct conversation, such as role-playing, unnecessary lists or sections, or formalities like greetings and sign-offs. However, it still maintains a tone that speaks directly to the student. There could be minor typos or grammar errors, but they don't affect the meaning or clarity. \\
    {\bf 0 points:} The feedback doesn't directly address the student, is overly long, or has major language issues that make it unclear or confusing. It may also misunderstand the task, be off-topic, or contain excessive repetition.
    \end{tcolorbox}
    \caption{Complete rubric description}
    \label{fig:rubric_detail}
\end{figure}

\subsection{Context: Explain in Plain English questions}
\label{sec:eipe}

We explore this rubric in the context of Explain in Plain English (EiPE) questions~\cite{murphy2012ability}, which present students with a piece of code and ask them to describe its functionality in plain English, as shown in Figure~\ref{fig:code_reading_question}.
These questions are typically designed to assess students' holistic understanding of the code, going beyond simple line-by-line descriptions.

We present six feedback examples and their corresponding human ratings in Table~\ref{tab:example}, covering the full range of scores for each rubric criterion. See Section~\ref{sec:human_feedback_collection} for details on the rating procedure.

\begin{table*}[!htbp]
    \caption{Examples of feedback with human ratings, drawn from the set of 360 feedback instances.}
    \includegraphics[clip, trim=2.5cm 6.7cm 4.2cm 3.3cm, width=1.00\textwidth]{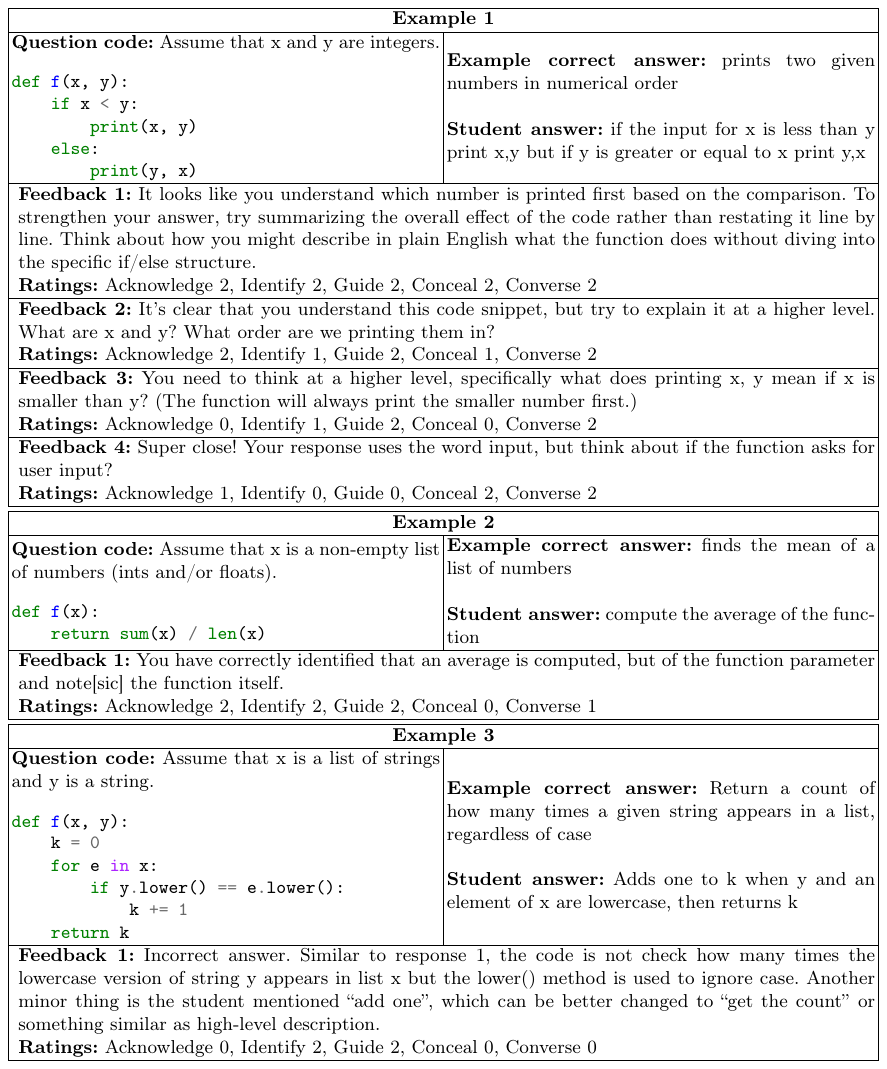}
    \label{tab:example}
\end{table*}

\subsection{Rubric development}

The rubric development process began with the first two authors experimenting with local models to generate feedback for EiPE questions. Through an iterative process of identifying issues in the generated feedback, grouping related issues into cohesive categories, and incorporating insights from instructor interviews and the literature on effective feedback, the research team drafted a rubric close to its final form. The first three authors made further refinements during the calibration and labeling process, which led to the finalized rubric presented in Figure~\ref{fig:rubric_detail}.

We initially considered a four-point scale (0–3) but found distinguishing that many levels for each criterion to be unreliable.  The trade-off between reliability and number of levels in an evaluation scale is well documented~\cite{wormeli2023fair, wolf2008tale}. In hindsight, our levels correspond to the ``there, somewhat there, not there yet'' scale~\cite{allen2006rubrics, bresciani2004assessing}.

\section{Data Collection}
\label{sec:data_collection}

With IRB permission, this study includes both actual student responses and TA-written feedback, in addition to LLM-generated feedback~\footnote{The entire dataset along with documentation can be found at \url{https://tinyurl.com/6ajvztr2}. Certain information has been anonymized for review.}

\subsection{Student responses} 
We used 15 EiPE questions and 90 student responses (six per question), selected from deidentified historical data to cover a range of issues that students encountered.

The EiPE questions and student responses used in this study are from an introductory Python programming course with a typical enrollment of approximately 300 students per semester at a large R1 US university. 

These questions have been used as formative and summative assessments for several years, but because of manpower issues, the course only routinely provides students with binary (correct/incorrect) feedback along with example correct answers. 

\subsection{Human feedback collection}
\label{sec:human_feedback_collection}

To collect human feedback that reflects realistic performance, we recruited nine course TAs to provide feedback on student responses. These TAs had varying levels of experience in the course; the number of semesters each TA served in the course is reported in Table~\ref{tab:ta_llm_stats}. The feedback collection was conducted at the end of the semester without time constraints or hard deadlines. TAs received no training on feedback provision specifically for this study and instead relied on their existing experience from interacting with students throughout the semester. This approach ensured that TAs had at least one full semester of experience in the course before providing feedback, while minimizing the risk of overfitting to our rubric and keeping their feedback as authentic as possible.

For human time constraints, the 90 student responses were randomly divided into three disjoint sets of 30 responses each, with each set containing two responses from each of the 15 questions. Each TA was randomly assigned to complete one set via a Google survey, ensuring that each set had feedback written for it by three different TAs. Instructions containing guidelines corresponding to each of the five rubric criteria were provided to the TAs at the beginning of the survey, as shown in Figure~\ref{fig:prompt_ta_feedback_a}. The format of each survey item is shown in Figure~\ref{fig:prompt_ta_feedback_b}.

\begin{table*}[t]
    \centering
    \caption{Mean human ratings and potential contributing factors for TA and LLM feedback.}
    \resizebox{\columnwidth}{!}{
        \begin{tabular}{c|cccccccccc}
                                       & TA 1 & TA 2 & TA 3 & TA 4 & TA 5 & TA 6 & TA 7 & TA 8 & TA 9 & LLM  \\ \hline
           Mean human rating           & 8.23 & 8.10 & 7.77 & 8.67 & 8.00 & 7.03 & 9.63 & 7.80 & 5.97 & 8.50 \\
           Mean feedback word count    & 22.4 & 29.4 & 19.7 & 37.4 & 20.7 & 22.0 & 64.6 & 83.2 & 17.5 & 60.9 \\
           Semesters of experience     & 1    & 1    & 4    & 1    & 1    & 5    & 4    & 1    & 2    & N/A
        \end{tabular}
    }
    \label{tab:ta_llm_stats}
\end{table*}

\begin{figure}[!ht]
    \centering
    \subcaptionbox{Instruction at the beginning of survey\label{fig:prompt_ta_feedback_a}}{
    \begin{tcolorbox}[width=\columnwidth, colback=gray!10, colframe=black]
    As a TA for CS 105, you’re likely familiar with the type of questions we assign where students analyze a piece of code and explain its functionality in plain English. These code reading questions are currently autograded on homework, providing binary correct/incorrect feedback. While efficient, this approach falls short of helping students improve their understanding or answers. \\

    To address this, we are exploring the use of large language models to deliver more nuanced, natural-language feedback. However, to develop and evaluate this system effectively, we need to understand how human-generated feedback compares. Specifically, we want to gather data on how TAs like you would write feedback in an ideal scenario where resources for detailed, hand-graded responses are unlimited. \\

    This form contains 30 student responses for you to review. Each page will present a code reading question, along with two student responses. The code itself, assumptions about function arguments, and example correct answers will also be provided for context. Your task is to provide constructive feedback to student for each response, following these guidelines: \\
    
    \begin{enumerate}
        \item Give actionable suggestions: Offer specific advice that could help the student better understand the code or improve their answer.
        \item Recognize correct elements: Highlight parts of the code the student has understood correctly, if any.
        \item Identify key mistakes: Point out significant errors or misunderstandings in the student's response, if any.
        \item Avoid revealing the answer: Do not directly provide the correct solution.
        \item Speak directly to the student. Write your feedback as if addressing the student personally, rather than discussing their response in third-person terms.
    \end{enumerate}

    Approach this as if you were grading and providing feedback on hand-graded homework questions. \\

    For your convenience, if you’d like to revisit these instructions during the process, here is a pastebin link with the full text: \href{https://pastebin.com/YDrM01ce}{https://pastebin.com/YDrM01ce} Feel free to open it in a separate tab for reference. \\

    Thank you for your time and effort in helping us improve feedback mechanisms for our students!
    \end{tcolorbox}
    }
    \caption{Prompt for TA (continues on next page)}
\end{figure}

\begin{figure}[!ht]\ContinuedFloat
    \centering
    \subcaptionbox{Each survey item\label{fig:prompt_ta_feedback_b}}{
    \begin{tcolorbox}[width=\columnwidth, colback=gray!10, colframe=black]
    Assumption:\\
    Assume that <assumption>.\\
    
    Code:\\
    <code>\\
    
    Example correct answers:\\
    <example\_correct\_answers>\\
    
    Student response:\\
    <student\_response>\\
    
    Your feedback to student response:
    \end{tcolorbox}
    }
    \caption{Prompt for TA. Tokens with <> will be replaced with text related to the student response for each survey item.}
    \label{fig:prompt_ta_feedback}
\end{figure}

\subsection{LLM feedback collection}
\label{sec:llm_feedback_collection}

To collect LLM feedback, we evaluated three models on a subset of 15 student responses: OpenAI o1 (via the ChatGPT Plus web interface with memory disabled), OpenAI o1-preview (via API), and DeepSeek-R1 (via web interface). Based on this evaluation, we selected OpenAI o1 (web interface) for the full study due to its superior performance. At the time of feedback collection, API access for OpenAI o1 was not available, necessitating the use of the web interface. The prompt template used to obtain LLM feedback is shown in Figure~\ref{fig:prompt_llm_feedback}. The LLM received effectively the same instructions and student response information as the TAs, including guidelines corresponding to each of the five rubric criteria. The slight difference in guidelines occurred because we collected human feedback first, then refined and finalized the rubric before collecting LLM feedback.

\begin{figure}[!ht]
    \centering
    \begin{tcolorbox}[width=\columnwidth, colback=gray!10, colframe=black]
    Code reading questions are questions where students are given a short piece of code, and asked to write a short, high-level English language description of what the code does. These questions assess students' ability to understand the functionality of the code and communicate the code's purpose rather than just the understanding of the syntax. Thus a line-by-line description is typically not accepted as correct. \\
    
    You are tasked to provide feedback to students on code reading questions with the following guidelines (your feedback doesn't need to address them in the order given below):
    \begin{enumerate}
        \item Offer a specific suggestion that could help the student better understand the code or improve their answer.
        \item Highlight parts of the code the student has understood correctly, if any.
        \item Point out significant errors, misunderstandings or omissions of important details in the student's response, if any.
        \item Do not give away the correct answer or reveal parts of the correct answer beyond what the student already understands.
        \item Write your feedback as if addressing the student personally, rather than discussing their response in third-person terms, keep it a few sentences long.
    \end{enumerate}
    A student is asked to describe the following piece of code with the assumption that <assumption>: \\
    
    <code> \\
    
    Example correct answers: \\
    <example\_correct\_answers> \\

    The student answered: \\
    <student\_response> \\

    Your feedback:
    \end{tcolorbox}
    \caption{Prompt template for LLM feedback. Tokens with <> will be replaced with text related to the student response to finalize the prompt.}
    \label{fig:prompt_llm_feedback}
\end{figure}

\section{Evaluation Process}
\label{sec:evaluation_process}

\subsection{Human rating process}
\label{sec:human_rating_process}

The first three authors conducted the human rating process. All had prior experience with projects related to EiPE questions and were familiar with this question type. The first author had served as a teaching assistant for the course, while the third author had been the course instructor for multiple semesters, giving both extensive experience interacting with students in this context. In contrast, the second author had no prior experience with students in this specific course. They are referred to as Raters 1–3 hereafter.

The raters conducted three rounds of calibration before rating the remaining feedback. In each calibration round, a sample of 20 feedback instances---equally split between TA-generated and LLM-generated feedback---was independently scored by the three raters. They then met to reconcile disagreements and refine the rubric. After completing three rounds of calibration, the raters independently scored the remaining feedback without further reconciliation. Throughout both calibration and rating, the raters were blinded to whether the feedback was TA-generated or LLM-generated but had access to the student response and question information.

After all feedback was scored by the three raters, we used the median of the three scores as the final rating for each criterion. The overall score was calculated as the sum of these median criterion scores. Throughout this paper, ``human ratings'' refers to these median ratings unless we explicitly reference individual rater scores (e.g., when reporting inter-rater reliability).

\subsection{LLM rating process}
\label{sec:llm_rating_collection}

To collect LLM ratings of feedback, we evaluated three models on a subset of 15 feedback instances: OpenAI o1, OpenAI GPT-4o, and OpenAI o3-mini (high reasoning), all accessed via the ChatGPT Plus web interface with memory disabled. Based on this evaluation, we selected OpenAI o1 due to its superior performance. We then obtained LLM ratings for all TA and LLM feedback using the OpenAI o1 API. The prompt template used to obtain LLM ratings is shown in Figure~\ref{fig:prompt_llm_evaluation}. The LLM was provided with the detailed rubric shown in Figure~\ref{fig:rubric_detail}. Compared to human raters, the LLM received the same rubric but lacked the calibration experience.

\begin{figure}[!ht]
    \centering
    \begin{tcolorbox}[width=\columnwidth, colback=gray!10, colframe=black]
    Code reading questions are questions where students are given a short piece of code, and asked to write a short, high-level English language description of what the code does. These questions assess students' ability to understand the functionality of the code and communicate the code's purpose rather than just the understanding of the syntax. Thus a line-by-line description is typically not accepted as correct.
    
    You are tasked to evaluate feedback provided to students on code reading questions with the following five rubric items: \\

    \{Complete rubric detail text in Figure~\ref{fig:rubric_detail}\} \\
    
    A student is asked to describe the following piece of code with the assumption that <assumption>: \\
    
    <code> \\
    
    Example correct answers: \\
    <example\_correct\_answers> \\
    
    The student answered: \\
    <student\_response> \\
    
    Feedback: \\
    <feedback> \\
    
    Your scoring on each rubric item (just give five integers separated by a single space):
    \end{tcolorbox}
    \caption{Prompt template for LLM evaluation. Tokens with <> will be replaced with text related to the feedback to finalize the prompt.}
    \label{fig:prompt_llm_evaluation}
\end{figure}

\section{Results}
\label{sec:results}

\subsection{Inter-rater reliability}
\label{sec:irr}

To assess inter-rater reliability, we report quadratic weighted Cohen's kappa between each pair of raters in Table~\ref{tab:cohen} and the distribution of score differences between raters in Table~\ref{tab:disagreement}.

Among human raters, almost all kappa values exceed 0.4, indicating moderate agreement, while Conversationality has two pairings between 0.2 and 0.4, reflecting fair agreement~\cite{landis1977measurement}. Most human rater pairs exhibit similar kappa values across criteria, except for Conversationality, where the Rater 1--2 and Rater 2--3 pairs show lower kappa values compared to the Rater 1--3 pair.

The primary reason for this difference is that most feedback received a Conversationality score of 2, which all human raters consistently recognized. Consequently, even a small number of disagreements resulted in noticeable kappa differences. 
The distribution of score differences supports this explanation, showing that human raters agreed on Conversationality 90\% of the time, while the Rater 1--2 and Rater 2--3 pairs had a slightly higher proportion of score differences of 1 than the Rater 1--3 pair.

Compared to agreement among human raters, the agreement between human raters and the LLM is substantially lower. This is evident from kappa values that are generally 0.1 lower and from the difference in mean agreement rates: around 0.8 among human raters versus around 0.7 between human raters and the LLM.



\begin{table*}[t]
    \centering
    \caption{Quadratic weighted Cohen's kappa for all 360 feedback instances, measured between each pair of raters. The first five columns correspond to individual rubric criteria, while the \textbf{Overall} column corresponds to the sum of all criteria scores for each feedback instance.}
    \resizebox{\columnwidth}{!}{
        \begin{tabular}{c|ccccc|c}
           Raters & Acknowledgment & Identification & Guidance & Concealment & Conversationality & Overall \\ \hline
           Rater 1 \& Rater 2 & 0.521 & 0.458 & 0.518 & 0.724 & 0.238 & 0.550 \\
           Rater 1 \& Rater 3 & 0.608 & 0.440 & 0.573 & 0.673 & 0.457 & 0.582 \\
           Rater 2 \& Rater 3 & 0.521 & 0.461 & 0.602 & 0.746 & 0.263 & 0.548 \\ \hline
           Rater 1 \& LLM     & 0.400 & 0.229 & 0.340 & 0.627 & 0.288 & 0.414 \\
           Rater 2 \& LLM     & 0.334 & 0.280 & 0.255 & 0.549 & 0.288 & 0.305 \\
           Rater 3 \& LLM     & 0.288 & 0.142 & 0.415 & 0.525 & 0.259 & 0.310 \\
        \end{tabular}
    }
    \label{tab:cohen}
\end{table*}


\begin{table*}[t]
    \centering
    \caption{Distribution of score differences across all 360 feedback instances between each pair of raters. Each cell shows the fraction of feedback instances with the corresponding score difference between the two raters. A score difference of 0 indicates agreement. The \textbf{Mean} row represents the average of each column.}
    \label{tab:disagreement}
    \resizebox{\columnwidth}{!}{
        \begin{tabular}{c|ccc|ccc|ccc}
        Raters            & \multicolumn{3}{c|}{Rater 1 \& Rater 2}                                     & \multicolumn{3}{c}{Rater 1 \& Rater 3}                                     & \multicolumn{3}{|c}{Rater 2 \& Rater 3}                                     \\ \hline
        Score difference  & \multicolumn{1}{c}{0}     & \multicolumn{1}{c}{1}     & 2     & \multicolumn{1}{c}{0}     & \multicolumn{1}{c}{1}     & 2     & \multicolumn{1}{c}{0}     & \multicolumn{1}{c}{1}     & 2     \\ \hline
        Acknowledgment    & \, 0.714 \, & \, 0.106 \, & \, 0.181 \, & \, 0.781 \, & \, 0.081 \, & \, 0.139 \, & \, 0.744 \, & \, 0.047 \, & \, 0.208 \, \\
        Identification    & 0.672 & 0.267 & 0.061 & 0.717 & 0.214 & 0.069 & 0.686 & 0.242 & 0.072 \\
        Guidance          & 0.786 & 0.164 & 0.050 & 0.806 & 0.153 & 0.042 & 0.828 & 0.128 & 0.044 \\
        Concealment       & 0.800 & 0.161 & 0.039 & 0.767 & 0.186 & 0.047 & 0.781 & 0.192 & 0.028 \\
        Conversationality & 0.939 & 0.056 & 0.006 & 0.931 & 0.067 & 0.003 & 0.919 & 0.072 & 0.008 \\ \hline
        Mean              & 0.782 & 0.151 & 0.067 & 0.800 & 0.140 & 0.060 & 0.792 & 0.136 & 0.072 \\
        \end{tabular}
    }
    \par\vspace{2.5mm}
    \resizebox{\columnwidth}{!}{
        \begin{tabular}{c|ccc|ccc|ccc}
        Raters            & \multicolumn{3}{c|}{Rater 1 \& LLM}                                     & \multicolumn{3}{c}{Rater 2 \& LLM}                                     & \multicolumn{3}{|c}{Rater 3 \& LLM}                                     \\ \hline
        Score difference  & \multicolumn{1}{c}{0}     & \multicolumn{1}{c}{1}     & 2     & \multicolumn{1}{c}{0}     & \multicolumn{1}{c}{1}     & 2     & \multicolumn{1}{c}{0}     & \multicolumn{1}{c}{1}     & 2     \\ \hline
        Acknowledgment    & \, 0.600 \, & \, 0.236 \, & \, 0.164 \, & \, 0.575 \, & \, 0.192 \, & \, 0.233 \, & \, 0.561 \, & \, 0.200 \, & \, 0.239 \, \\
        Identification    & 0.547 & 0.408 & 0.044 & 0.581 & 0.353 & 0.067 & 0.569 & 0.356 & 0.075 \\
        Guidance          & 0.633 & 0.314 & 0.053 & 0.611 & 0.311 & 0.078 & 0.672 & 0.278 & 0.050 \\
        Concealment       & 0.794 & 0.150 & 0.056 & 0.750 & 0.172 & 0.078 & 0.694 & 0.236 & 0.069 \\
        Conversationality & 0.939 & 0.061 & 0.000 & 0.947 & 0.050 & 0.003 & 0.914 & 0.083 & 0.003 \\ \hline
        Mean              & 0.703 & 0.234 & 0.063 & 0.693 & 0.216 & 0.092 & 0.682 & 0.231 & 0.087 \\
        \end{tabular}
    }
\end{table*}

\subsection{Feedback quality: TA vs LLM}

\begin{figure}[!ht]
    \centering
    \includegraphics[width=0.89\columnwidth]{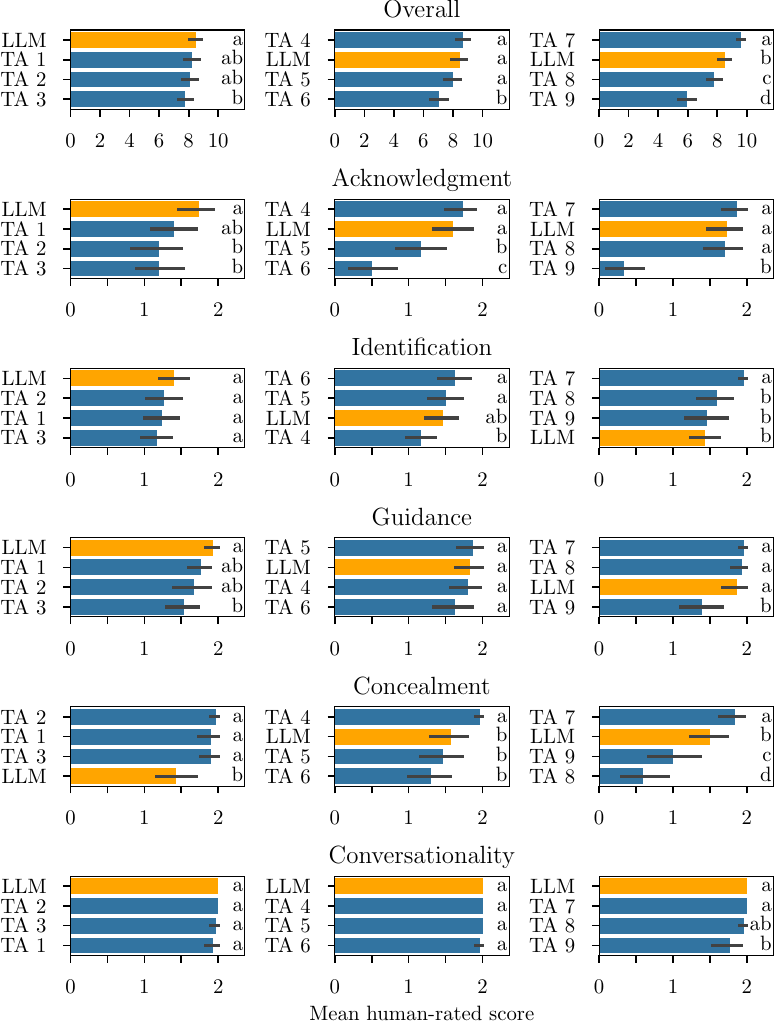}
    \vspace{5pt}
    \caption{Mean scores for each TA and LLM on overall score and individual rubric criteria, from human raters. Each column represents one of the three sets used during human feedback collection. Error bars represent 95\% confidence intervals. Compact letter display~\cite{piepho2004algorithm} shows results of pairwise comparisons using paired t-tests without multiple hypothesis correction; bars sharing the same letter are not statistically significantly different.}
    \label{fig:barplot_mean_score}
\end{figure}

To compare the quality of feedback provided by TAs and the LLM, we plotted the mean score of each TA and the LLM as evaluated by human raters in Figure~\ref{fig:barplot_mean_score}. As shown in the figure, the LLM consistently ranks among the top two performers in most cases. When it falls into the bottom two, it remains close behind the leaders. Only two TAs outperformed the LLM in terms of overall score, while some TAs performed markedly worse.

\begin{figure}[!ht]
    \centering
    \includegraphics[width=0.99\columnwidth]{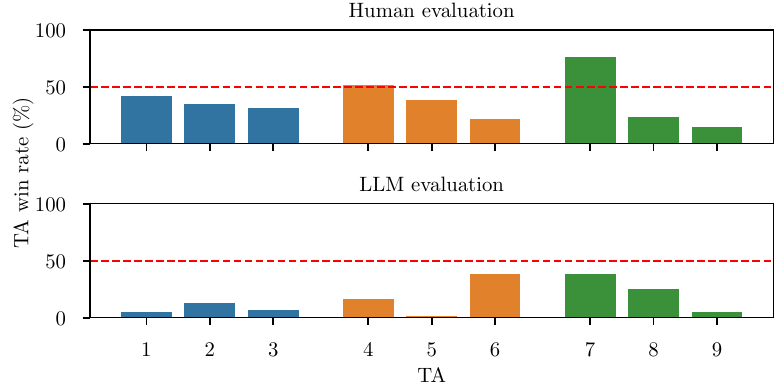}
    \vspace{5pt}
    \caption{TA win rate against the LLM in feedback quality for the set of student responses each TA reviewed. TAs reviewing the same set of student responses are grouped by color. The top plot shows results from human evaluations; the bottom plot shows results from LLM evaluations.}
    \label{fig:ta_win_rate}
\end{figure}

To assess whether TAs are consistently better or worse than the LLM, we plotted the TA win rate in Figure~\ref{fig:ta_win_rate}. The TA win rate represents the fraction of feedback instances where a TA received a higher overall score than the LLM, with ties split evenly (half a point each). As shown in the figure, every TA has, at times, provided better feedback than the LLM, as judged by human raters. Even TA 9, who appears markedly worse than the LLM based on mean overall score, has occasionally produced higher-quality feedback. Similarly, TAs 4 and 7, despite outperforming the LLM in terms of mean overall score, do not win 100\% of the time.

\begin{figure}[!ht]
    \centering
    \includegraphics[width=0.99\columnwidth]{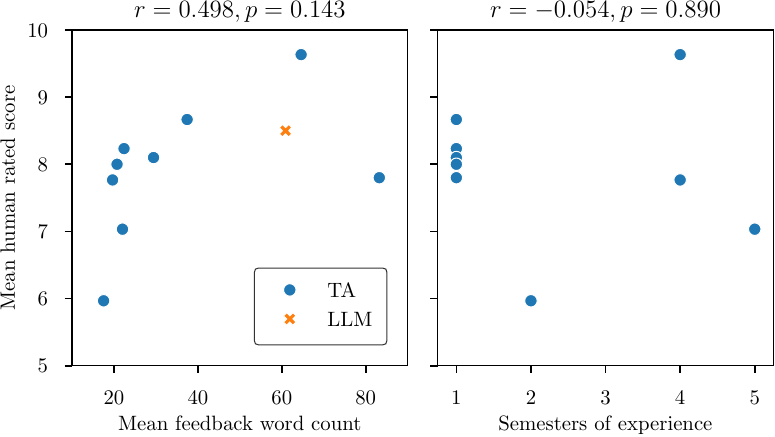}
    \vspace{5pt}
    \caption{Mean human ratings plotted against mean feedback word count (left) and semesters of TA experience (right). Note: axes do not start at zero. This visualizes data from Table~\ref{tab:ta_llm_stats}.}
    \label{fig:scatterplot_of_mean_human_rating}
\end{figure}

We examined potential factors that might influence feedback quality. Specifically, we investigated TA experience (measured by the number of semesters served in the course) and feedback length (measured by word count). These metrics, along with mean human ratings, are reported in Table~\ref{tab:ta_llm_stats} and visualized in Figure~\ref{fig:scatterplot_of_mean_human_rating}. The correlation between mean human rating and mean feedback word count is mildly positive (r = 0.498, p = 0.143, including the LLM), suggesting that longer feedback tends to receive higher human ratings. In contrast, the correlation between mean human rating and TA experience is near zero (r = -0.054, p = 0.890).


\subsection{Consistency/variance: TA vs LLM}

\begin{figure}[!ht]
    \centering
    \includegraphics[width=0.99\columnwidth]{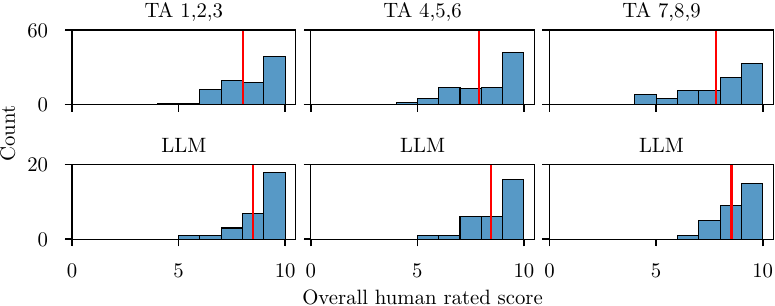}
    \vspace{5pt}
    \caption{Distribution of overall scores for TA and LLM feedback, as evaluated by human raters. Each column represents one of the three sets of student responses that TAs reviewed. TAs who wrote feedback the same set are grouped together. The y-axis is scaled so that TA counts are three times those of the LLM (matching the 3:1 ratio of TAs to LLM), allowing direct visual comparison of distribution shapes. Red lines indicate the mean of each distribution.}
    \label{fig:histplot_overall_score}
\end{figure}

To examine the consistency and variance in feedback quality, we plotted the distribution of overall scores for TAs and the LLM as evaluated by human raters in Figure~\ref{fig:histplot_overall_score}. The TAs are grouped such that each column corresponds to those who reviewed the same set of student responses. Since each column includes feedback from three TAs compared to one LLM, the y-axis ticks are scaled so that TA counts are three times those of the LLM, enabling direct visual comparison of distribution shapes and spreads. As shown in the figure, TA feedback exhibits a wider spread and lower mean overall score compared to the LLM. Specifically, the LLM more consistently provides higher-quality feedback: 79\% of LLM feedback scores 8 or above, whereas only 62\% of TA feedback reaches this threshold.

\subsection{LLM rating bias}

\begin{figure}[!ht]
    \centering
    \includegraphics[width=0.99\columnwidth]{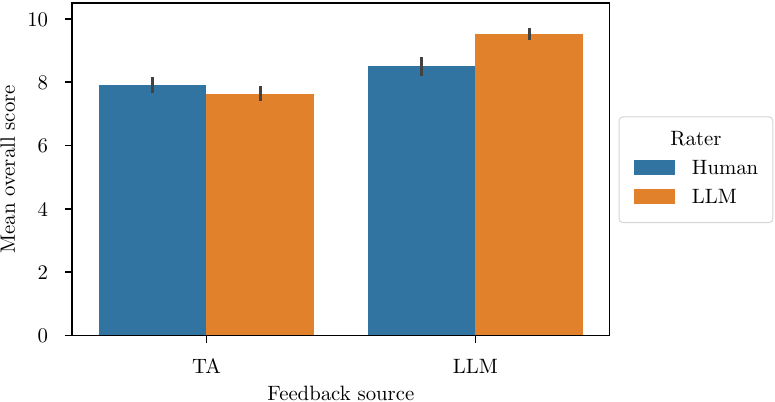}
    \vspace{5pt}
    \caption{Mean overall scores for TA and LLM feedback as evaluated by human raters versus by LLM. Error bars represent 95\% confidence intervals. All groups are statistically significantly different without correction for multiple hypothesis testing. Paired t-tests were used for comparisons within the same feedback source; Welch's t-tests were used for comparisons between different feedback sources.}
    \label{fig:llm_bias_overall}
\end{figure}

\begin{figure}[!ht]
    \centering
    \includegraphics[width=0.99\columnwidth]{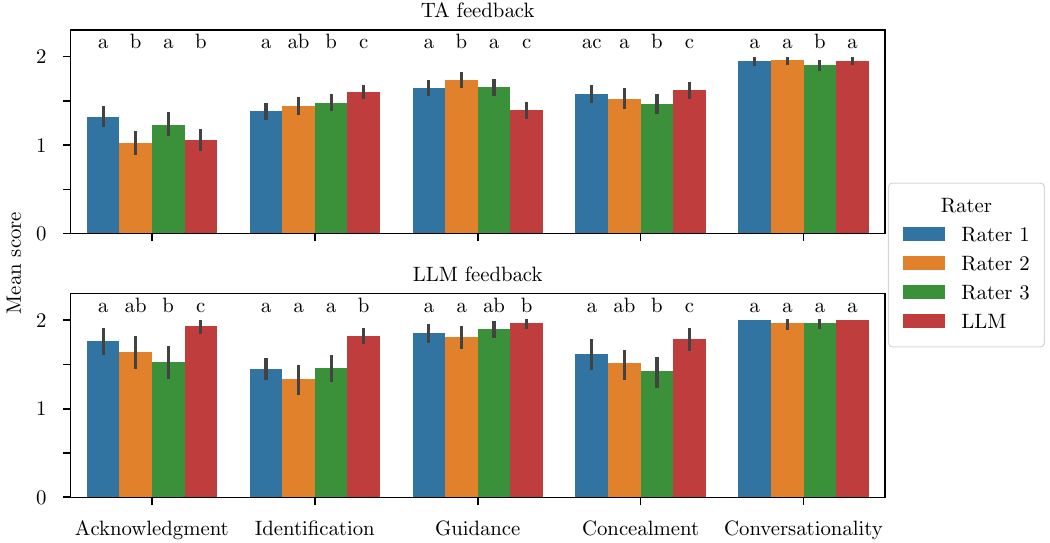}
    \vspace{5pt}
    \caption{Mean scores for TA and LLM feedback across all rubric criteria, evaluated by each rater. Error bars represent 95\% confidence intervals. Compact letter display~\cite{piepho2004algorithm} shows results of pairwise paired t-tests for each rubric criterion within each feedback source, without correction for multiple hypothesis testing. Bars sharing the same letter are not statistically significantly different.}
    \label{fig:llm_bias_detail}
\end{figure}

To examine whether the LLM can serve as a reliable evaluation agent, we investigated whether the LLM exhibits bias when rating TA versus LLM feedback. We plotted the mean overall scores under conditions where human raters or the LLM evaluated only TA feedback or only LLM feedback in Figure~\ref{fig:llm_bias_overall}. As shown in the figure, the LLM assigns lower ratings to TA feedback compared to human raters and higher ratings to LLM feedback compared to human raters, suggesting that the LLM is biased toward feedback it generates. This trend is also reflected in Figure~\ref{fig:ta_win_rate}, where human evaluation generally yields higher TA win rates compared to LLM evaluation. A detailed breakdown across all rubric criteria and individual raters is shown in Figure~\ref{fig:llm_bias_detail}. As the figure shows, the LLM tends to underrate TA feedback on Acknowledgment and Guidance while overrating LLM feedback on Acknowledgment, Identification, and Concealment.

\section{Discussion}
\label{sec:discussion}

\subsection{RQ 1: Reflections on the rubric}

Based on our experience applying the rubric, we found that it effectively systematized the evaluation of feedback quality. The scores aligned with our intuitions about feedback quality, and we did not feel the absence of important criteria for our context. We believe the rubric successfully captures the essential dimensions of effective feedback for code reading tasks.

We propose that three of our criteria---Acknowledgment, Identification, and Guidance---constitute the core pillars of effective feedback that likely generalize across domains. Acknowledgment ensures students understand what they did correctly, preventing them from inadvertently degrading their answers and supporting their confidence. Identification clarifies where the response falls short, addressing misconceptions or omissions. Guidance provides actionable direction for improvement, answering Hattie's ``Where to next?'' question~\cite{hattie2007power}. Together, these three criteria form a foundation for balanced, constructive feedback regardless of subject matter.

The remaining criteria---Concealment and Conversationality---appear more context-dependent and may warrant modification depending on the learning environment. For instance, we did not include a ``feed-up'' criterion addressing Hattie's ``Where am I going?'' question~\cite{hattie2007power}, as the learning goal in code reading tasks is clear: describe what the code does at a high level. However, such a criterion might be essential in contexts like essay writing, where reminding students of the assignment's objectives and evaluation standards could meaningfully guide their revisions. Similarly, the importance of Concealment may vary: while withholding correct answers on initial attempts encourages independent reasoning, providing more direct information after multiple unsuccessful attempts may better serve struggling students.


Our experience revealed several practical considerations for rubric refinement. First, we observed that the qualitative distinction between scores of 0 and 1 for each criterion was substantially greater than between 1 and 2. A score of 0 indicated a fundamental absence or failure of that feedback criterion, whereas scores of 1 and 2 both represented functional feedback with varying degrees of completeness or sophistication. Since the bulk of ratings---for both human and LLM feedback---fell in the 1--2 range, finer granularity at this end of the scale provides better resolution.  In practice, feedback scoring 8–10 overall was generally acceptable, while scoring 0 in any single criterion rendered the feedback unacceptable.

Second, equal weighting of criteria may not reflect their relative importance, at least in the 1--2 range. Scoring a 1 versus 2 in Identification or Conversationality seemed less consequential than in other criteria. Explicitly identifying an error appears less critical when clear, concrete guidance for improvement is provided. Similarly, minor typos (Conversationality) are unlikely to undermine feedback usefulness. In fact, Conversationality may be a candidate for removal entirely: nearly all feedback in our study received a rating of 2, with only TA 9 scoring lower (Figure~\ref{fig:barplot_mean_score}). This criterion emerged during rubric development when smaller local models sometimes failed to produce coherent feedback---going off-task, using incorrect tone, or addressing the wrong audience. Frontier models rarely exhibit such failures, suggesting this criterion may no longer be essential.

In contrast, Guidance likely warrants the highest weight, as it directly provides actionable advice for students. For feedback on initial attempts, Concealment may be second in importance, as it encourages independent problem-solving. However, its importance diminishes---or even reverses---for subsequent attempts when students are struggling. Acknowledgment ranks third in importance: reinforcing correct elements prevents students from degrading their answers and supports their motivation and confidence.

These reflections suggest that while our rubric effectively captured feedback quality in our context, adaptations would improve its utility. Future work might explore weighted scoring schemes, context-specific criterion modification, and validation across diverse question types and domains to establish which criteria remain essential and which require tailoring.

\subsection{RQ 2: Comparison of LLM and TA feedback quality}
\label{sec:discusssion_feedback_quality}

Our evaluation demonstrates that the LLM performs on par with or above the average TA in our sample. This finding has significant practical implications: in large courses, TAs are typically responsible for day-to-day student interactions and feedback provision. The fact that LLMs can match or exceed typical TA performance suggests substantial potential for real-world impact, even before considering their unique operational advantages.

Indeed, LLMs possess several strengths that human instructors simply cannot parallel: 24/7 availability, response times measured in seconds rather than hours or days, ability to scale to unlimited numbers of students simultaneously, and most notably, consistent performance across all interactions. Our data clearly illustrate this last advantage---while TA feedback quality varied substantially, the LLM output maintained consistent quality (Figure~\ref{fig:barplot_mean_score} and \ref{fig:histplot_overall_score}). This consistency may be particularly valuable in educational settings where students benefit from predictable and reliable feedback experiences.

These operational characteristics suggest that LLMs need not necessarily match the best human performance to generate a meaningful educational impact. A system providing consistently adequate feedback at scale and with immediate availability might well produce superior learning outcomes compared to higher-quality but limited human feedback. We anticipate that if a randomized controlled trial were conducted---not necessarily blinded, as constraining LLM response speed would be ecologically invalid---positive results would emerge. The quality parity we observe here strengthens confidence in any positive randomized controlled trial outcomes, as it establishes that improvements cannot be attributed solely to increased feedback frequency but reflect genuinely useful feedback content.

\subsection{RQ 2: Role of feedback length in quality}
\label{sec:discusssion_feedback_length}

An important methodological consideration emerges from examining prior work reporting LLM superiority over human instructors. In several such studies, the human feedback examples shown are substantially shorter than LLM-generated feedback~\cite{dai2024assessing, wan2024exploring}, thus often lacking the elaboration that Shute argues is an important feature of effective feedback~\cite{shute2008focus}. We suspect that the length of the feedback may be a primary contributor to the observed quality differences in those comparisons. Although one could argue that human instructors' inability to provide elaborated feedback at scale is precisely why LLMs offer practical advantages---a valid point for randomized controlled trial evaluations---we believe quality evaluation should provide both parties equal footing by accounting for this factor.

Fortunately, our sample included TAs with varying experience levels and feedback length preferences, allowing us to examine this relationship empirically. Our data confirm that length is indeed a contributing factor, as evidenced by the mild positive correlation between feedback length and quality scores (Figure~\ref{fig:scatterplot_of_mean_human_rating}). However, the relationship is not straightforward: longer feedback does not guarantee higher quality. TA 8 produced the longest feedback on average but scored below the LLM, while TA 1's relatively succinct feedback achieved scores nearly comparable to top performers (Table~\ref{tab:ta_llm_stats}). These findings suggest that while adequate length is necessary to address rubric criteria in a meaningful way, excessive length without corresponding quality may indicate verbosity rather than pedagogical value---consistent with Shute's conclusion that while elaboration is generally better than no elaboration, overly long or complicated feedback can overwhelm learners and reduce effectiveness~\cite{shute2008focus}.

\subsection{RQ 2: Sources of variance in TA performance}
\label{sec:discusssion_ta_variance}

Much of the variance in TA feedback quality appears to stem from differences in their ability or willingness to follow the provided directions. The highest-scoring TA (TA 7) followed a consistent format addressing all rubric criteria. Notably, this TA served as co-instructor for the course and possessed explicit interest and knowledge in pedagogy, which may have made the written guidance more comprehensible and memorable, translating to superior performance. Similarly, TA 4, the second highest-scoring TA, demonstrated exceptional engagement with teaching responsibilities, voluntarily creating multiple pedagogical resources for the course. This suggests that pedagogical knowledge and intrinsic motivation---rather than teaching experience alone---significantly influence feedback quality.

Lower-scoring TAs typically exhibited systematic deficiencies in specific rubric criteria rather than overall incompetence. TA 8's primary weakness was low Concealment scores; their feedback would have been appropriate for students after multiple attempts but revealed too much of the correct answer for initial attempts. TA 6 consistently under-performed on Acknowledgment, providing feedback focused entirely on what needed fixing without reinforcing correct elements. TA 9 scored poorly on both Acknowledgment and Concealment, typically offering feedback in the form ``Incorrect answer'' followed by the correct solution. This TA had extensive experience performing binary (correct/incorrect) grading for coding exercises in the course and may have approached the feedback task with a grading mindset rather than a formative learning perspective.

It should be emphasized that these TAs received no formal training on writing feedback messages beyond the written instructions provided at the beginning of the survey, which included five guidelines corresponding to the five rubric criteria. With targeted training, all TAs might achieve performance comparable to our best performers. However, it may not be ecologically valid to assume instructors routinely provide such training. In typical course settings, TAs often receive minimal pedagogical preparation, and their feedback quality may reflect this reality. This context further strengthens the case for LLM deployment: unlike human TAs, LLMs can be systematically prompted and fine-tuned to consistently follow best feedback practices without requiring ongoing training or supervision.

The variability in TA performance also raises questions about baseline comparisons in educational technology research. Studies comparing innovations against ``typical instructor feedback'' must grapple with the fact that instructor quality varies substantially. Our findings with TAs suggest that demonstrating LLM parity with average performance---rather than exceptional instructors---may be a more appropriate and practically relevant benchmark, as average performance better represents what most students actually experience in their courses.

\subsection{RQ 3: LLM as evaluator of feedback quality}

The use of LLMs as evaluators of LLM-generated content has become increasingly prevalent in recent AI research~\cite{zheng2023judging, dubois2024length, liu2023g}. This approach offers compelling advantages: scalability, cost-effectiveness, and the ability to provide nuanced assessments that go beyond simple metrics. Within educational contexts, several studies have explored using LLMs as primary evaluators of feedback quality~\cite{scarlatos2024improving, koutcheme2024open, jia2024assessing, scarlatos2025training}, motivated by the promise of automated, low-cost quality assessment that could enable rapid iteration and large-scale evaluation.

However, our findings urge caution when employing LLMs as evaluators without careful examination of potential biases. We observed a systematic pattern: LLMs tend to assign higher scores to feedback they themselves generated compared to human expert ratings, while assigning lower scores to TA-generated feedback (Figure~\ref{fig:llm_bias_overall}). This self-preference bias---where models favor their own outputs---has been widely documented in the LLM literature~\cite{wataoka2024self, panickssery2024llm, stureborg2024large}.

Recent research reveals that this bias persists even when using a different LLM as the evaluator, suggesting the phenomenon is more accurately characterized as familiarity bias~\cite{wataoka2024self}. LLM evaluators systematically assign higher ratings to text with lower perplexity---text that appears more predictable based on the model's training distribution~\cite{wataoka2024self, stureborg2024large}. Since LLMs from similar families often share comparable training data and stylistic conventions, they produce outputs with mutually low perplexity. This explains why cross-model evaluation fails to eliminate bias: the judge favors text aligned with its own learned patterns, regardless of actual authorship~\cite{wataoka2024self}.

These findings have important methodological implications for researchers considering LLM-based evaluation of educational feedback. Using the same model to generate and evaluate feedback creates a confounded assessment where observed ``quality'' partially reflects familiarity bias rather than genuine pedagogical value. Even employing a different model as a judge may not resolve this issue if both models share similar training distributions or architectural characteristics. Moreover, even when LLM evaluation is only applied to human-generated feedback alone---thereby avoiding self-preference bias entirely---there could still be systematic deviations from human judgment. As shown in our case in Figure~\ref{fig:ta_win_rate}, human evaluators rated TA 7's feedback substantially higher than TA 6's, whereas the LLM judged them as roughly equivalent. Similarly, human evaluators considered TA 4's feedback preferable to TA 6's overall, but the LLM reached the opposite conclusion.

These discrepancies suggest that LLM evaluators could exhibit biases beyond mere self-preference, potentially reflecting differences in how humans and models weigh various aspects of feedback. Validation against human expert ratings on a representative subset of the data is therefore essential to quantify the magnitude and direction of bias. This validation enables researchers to either apply statistical corrections to LLM ratings or appropriately caveat findings with explicit acknowledgment of bias patterns.

Despite these concerns, complete abandonment of LLM-based evaluation would be premature. LLMs offer genuine advantages for large-scale assessment and can provide a valuable signal when their limitations are understood and systematically accounted for. However, our results demonstrate that LLM evaluators cannot yet serve as unbiased substitutes for human judgment in the evaluation of feedback quality. Researchers should approach LLM-based evaluation with full awareness of its limitations and design studies that explicitly account for systematic bias rather than treating LLM ratings as ground truth.

\subsection{No obvious hard questions or responses for giving feedback}

When we started this work, we imagined there might be questions or student responses that would be hard to write feedback for, but we did not find any. 
All student responses had at least one good feedback (a rating of 8+) and only two questions were devoid of poor feedback (no feedback with a rating $\leq 5$).  To an extent, this is good news, as it means all student responses can receive feedback that's high-quality and hopefully useful. 
Furthermore, we didn't find a correlation between question difficulty and the ease of writing feedback for it. Nor in general is it the case that certain pieces of code generated \textit{unusual} student responses that were harder to write feedback for.

\section{Limitations}

Our study has a few limitations that should be considered when interpreting these findings.

This study examines feedback from one introductory programming course at a single institution. Feedback quality patterns and LLM performance may vary across different educational contexts, student populations, course levels, and subject domains. While programming education shares common pedagogical principles with other STEM disciplines, the specific characteristics of code-based assignments may not fully generalize to subjects like mathematics, physics, or writing.

Our analysis of LLM evaluation bias focused on OpenAI o1. While research suggests LLM self-preference bias generalizes across models and persists even in cross-model evaluation, the magnitude of bias may vary. We did not systematically test multiple LLM evaluators or explore debiasing techniques. Additionally, LLM-generated feedback reflects the capabilities of models available at the time of data collection (Late Fall 2024). Rapid improvements in LLM capabilities mean that our findings may not generalize to newer model generations, though the fundamental bias patterns we identified may persist given their connection to training distribution similarities.

We evaluated individual feedback instances without examining how students actually engage with and respond to different sources of feedback. A randomized controlled trial comparing student learning outcomes when receiving LLM-generated versus TA-generated feedback would provide stronger evidence of real-world pedagogical effectiveness. Our quality ratings, while based on established pedagogical principles, represent expert assessments of potential value rather than demonstrated impact on student learning.

\section{Conclusions and future work}

This study compared LLM-generated and TA-generated feedback for open-ended short answer questions in a CS1 course. We found that LLM-generated feedback generally matched or exceeded TA feedback quality based on a rubric we developed that has grounding in education literature. Our analysis also revealed significant LLM self-preference bias in automated evaluation, where an LLM systematically rated its own outputs higher than human experts did.

For the EiPE questions in the CS1 course, we feel hint-oriented LLM feedback is ready for deployment in formative contexts. It enables immediate feedback delivery, which benefits student learning, while reducing TA workload on routine assessments and allowing TAs to focus on students who need additional support.

Several directions merit further investigation. While this work focused on generating feedback for first attempts with concealment, frontier LLMs likely have equal capability for providing subsequent feedback that progressively discloses solutions and explains differences between the student's submitted answer and a correct answer. More broadly, exploring multi-round interactions where students engage in dialogue with the feedback system could reveal how LLMs support iterative problem-solving. Randomized controlled trials comparing student learning outcomes across feedback modalities would provide direct evidence of pedagogical effectiveness. Investigating LLM feedback quality across different subject domains (mathematics, physics, writing) would test generalizability beyond programming.

The ease of access to LLM-generated help raises important questions about impacts on student metacognitive development. On one hand, readily available assistance might reduce students' development of self-regulated learning strategies and problem-solving persistence. On the other hand, with systems that allow conversational interaction with LLMs---beyond the submit-and-receive-feedback paradigm examined in this paper---students could develop metacognitive skills by practicing question formulation and help-seeking behaviors that were previously constrained by limited office hours. Longitudinal studies are needed to understand these complex dynamics and their implications for long-term learning outcomes.



%
%
%
\bibliographystyle{elsarticle-harv}
\bibliography{references}

\end{document}